УДК [004.057.5+331.55]::[378::004.9]

**Є.О. Модло**
*Криворізький металургійний інститут
Національної металургійної академії України*

**Ю.В. Єчкало, С.О. Семеріков, В.В. Ткачук**
*ДВНЗ «Криворізький національний університет»*

# ВИКОРИСТАННЯ ТЕХНОЛОГІЇ ДОПОВНЕНОЇ РЕАЛЬНОСТІ У МОБІЛЬНО ОРІЄНТОВАНОМУ СЕРЕДОВИЩІ НАВЧАННЯ ВНЗ

*Мета дослідження:* висвітлення особливостей використання технології доповненої реальності у мобільно орієнтованому середовищі навчання ВНЗ.

*Завдання дослідження:* визначити роль та місце технології доповненої реальності у мобільно орієнтованому середовищі навчання, а також можливості використання технології доповненої реальності у навчанні фізики.

*Об'єкт дослідження:* мобільно орієнтоване середовище навчання ВНЗ.

*Предмет дослідження:* технологія доповненої реальності у мобільно орієнтованому середовищі навчання ВНЗ.

*Використані методи дослідження:* теоретичні – аналіз науково-методичної літератури; емпіричні – навчання, спостереження за навчальним процесом.

*Результати дослідження.* На основі аналізу наукових публікацій визначено поняття доповненої реальності. Відмічено, що онлайн-експерименти засобами доповненої реальності надають студентам можливість спостерігати й описувати роботу реальних систем при зміні їхніх параметрів, а також частково замінити експериментальні установки об'єктами доповненої реальності. Розглянуто схему реалізації доповненої реальності. Окремо виділено можливості роботи з об'єктами доповненої реальності у навчанні фізики. Показано, що застосування засобів доповненої реальності надає можливість підвищити реалістичність дослідження; забезпечує емоційний та пізнавальний досвід, що сприяє залученню студентів до систематичного навчання; надає коректні відомості про установку в процесі експериментування; створює нові способи подання реальних об'єктів у процесі навчання.

***Ключові слова:*** *технологія доповненої реальності, мобільно орієнтоване середовище навчання ВНЗ, навчання фізики.*

📖 **Постановка проблеми.** Використання мобільних Інтернет-пристроїв розширює межі традиційного інформаційно-освітнього середовища ВНЗ до мобільно орієнтованого – відкритої багатовимірної педагогічної системи, що включає психолого-педагогічні умови, мобільні інформаційно-комунікаційні технології і засоби навчання, наукових досліджень та управління освітою, і забезпечує взаємодію, співпрацю, розвиток особистості викладачів і студентів у процесі вирішення освітніх та наукових завдань у будь-який час та у будь-якому місці [9, с. 24]. Одним із шляхів підвищення ефективності мобільно орієнтованого навчального середовища є застосування технології доповненої реальності, що надає можливість об'єднання реальних та віртуальних засобів навчання за допомогою мобільних Інтернет-пристроїв.

**Аналіз останніх досліджень**. Дослідження в області доповненої реальності проводили Ф. Кішіно, Т.П. Коделл, Д.В. Майзел, П. Мілгрем, А.Е. Сазерленд та інші науковці. У їхніх працях розглядалися проблеми таксономії, розробки та використання засобів доповненої реальності у навчальному процесі та у професійній діяльності. Зокрема, роботи Н. Гуаєль, Е. Гуінтерса, Х. Мартін-Гутьєрреса, Д. Перес-Лопеса, М.Т. Рестіво, Т. Різова, Ж.-М. Сьотата, О. Хьюга підтвердили позитивний ефект використання даної технології у навчанні та надали можливість визначити застосування технології доповненої реальності як один із найбільш перспективних напрямків підвищення ефективності процесу навчання у вищих навчальних закладах.

**Метою статті** є висвітлення особливостей використання технології доповненої реальності у мобільно орієнтованому середовищі навчання ВНЗ.

**Виклад основного матеріалу**. Мобільні Інтернет-пристрої реалізують концепцію мобільного навчання [11] – навчання незалежно від часу та місця [8; 9; 10]. Приклади використання мобільних пристроїв у навчанні фізики студентів інженерних спеціальностей подано на рис. 1.

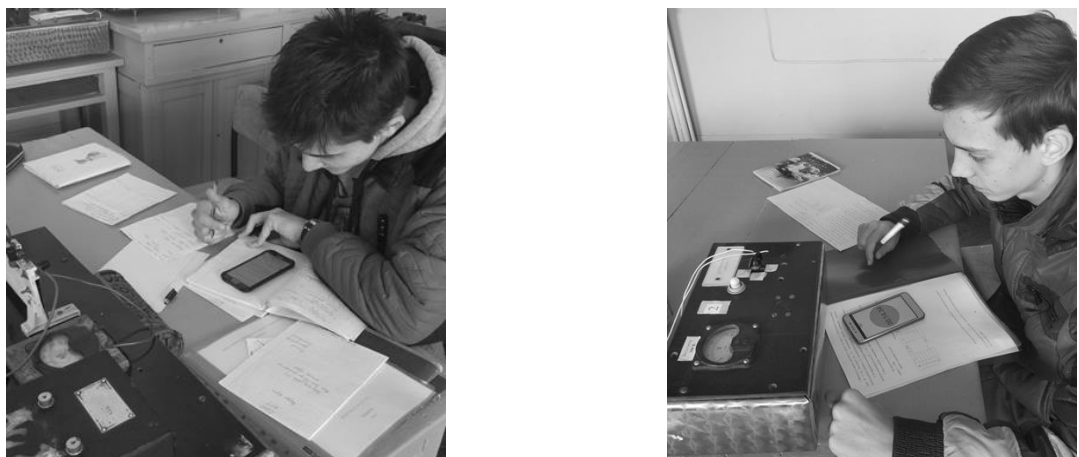

Рис. 1. Використання мобільних пристроїв на лабораторних роботах з фізики

Концепція віртуальної та доповненої реальності розвивається з 1960-х рр. і вважається дуже перспективним, потужним і корисним інструментом, особливо в освіті. Доповнена реальність визначається як поєднання фізичних та цифрових просторів у семантично пов'язаних контекстах, для яких об'єкти асоціацій розташовані у реальному світі [2]. На відміну від віртуальної реальності, доповнена не створює повністю віртуальне середовище, а поєднує віртуальні елементи з реальним світом: до реального оточення користувача додаються віртуальні об'єкти, що змінюються унаслідок його дій. «Батько» сучасних інтерфейсів користувача А. Е. Сазерленд у піонерській роботі 1968 року вказує, що це вимагає створення віртуальних інструментів або компонентів, керованих користувачем, для виконання певних дослідів, проведення експерименту тощо [7]. Розроблений ним шлем віртуальної та доповненої реальності має влучну назву «Дамоклів меч» – через велику вагу та розміри механізм був стаціонарно змонтований над користувачем (рис. 2).

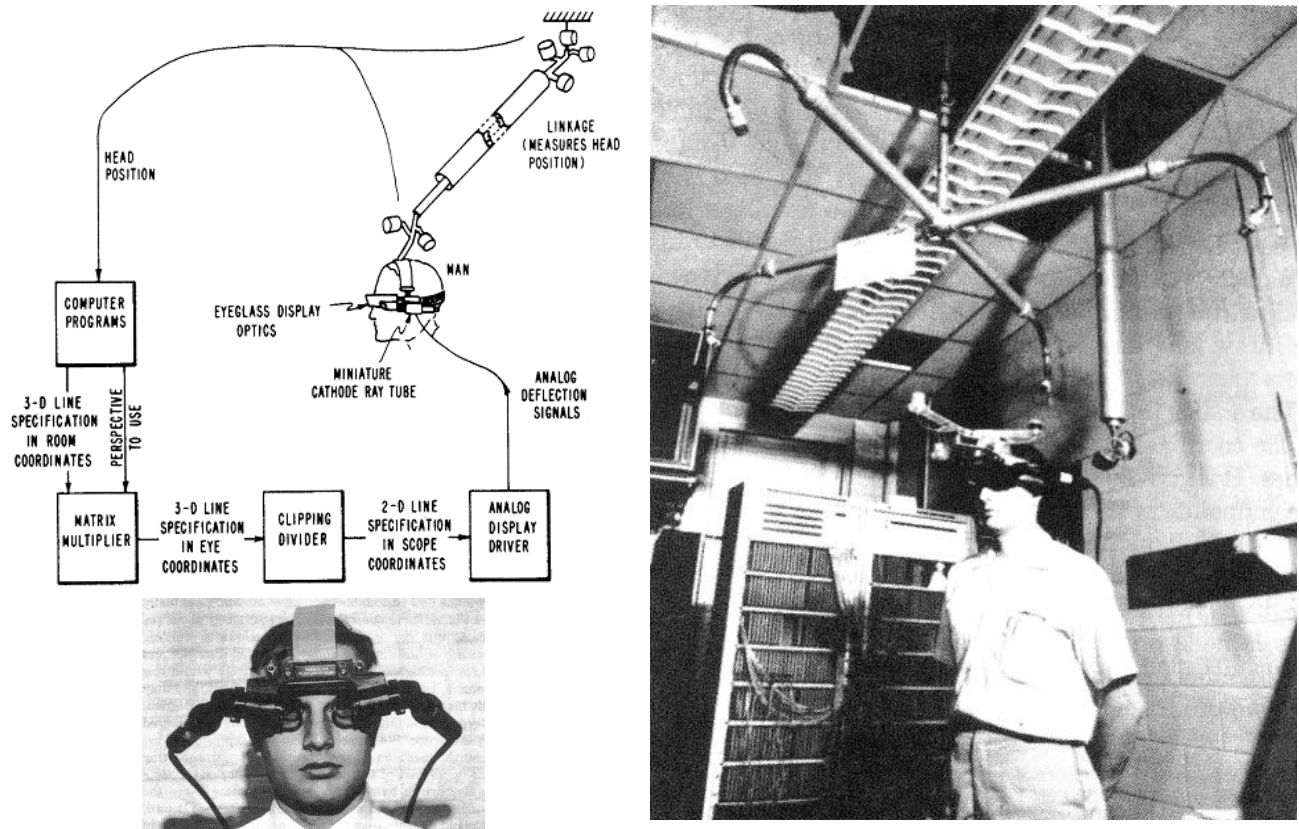

Рис. 2. Загальна схема роботи та зовнішній вигляд «Дамоклового меча» А.Е. Сазерленда та Р.Ф. Спрула [7, с. 296-298]

Таким чином, саме недостатня мобільність технології доповненої реальності стала основною перешкодою для її поширення – більше 30 років дослідження у цій галузі не виходили за межі окремих лабораторій. Теоретичним підсумком цього етапу стала робота П. Мілгрема та Ф. Кішіно: автори описують простір між реальним та віртуальним світом (називаючи його комбінованою реальністю), у якому доповнена реальність є більш близькою до реального (немодельованого) світу, а доповнена віртуальність – до віртуального (повністю модельованого) світу [4].

Лише з появою мобільних пристроїв у 1990-х рр. виникли технологічні передумови для використання технології доповненої реальності поза межами спеціалізованих лабораторій – у мобільному просторі Інтернет-користувача. На основі технології доповненої реальності були створені мобільні програмні засоби, призначені для вивчення різних дисциплін (соціально-гуманітарних, фундаментальних та фахових). За допомогою таких засобів надаються відомості про навчальні об'єкти та їхні характеристики. У ряді проектів, реалізованих у Північній Америці та Європі, мобільні пристрої використовувалися для візуалізації віртуальних об'єктів доповненої реальності. Так, за допомогою програмного забезпечення для мобільних пристроїв майбутні інженери могли бачити, де розташовуються опори мостів при їх візуальному огляді під різними кутами [2; 3].

М. Т. Рестіво та іншими авторами [5] було розглянуто можливості застосування технології доповненої реальності у навчанні розділу «Електрика» курсу фізики. Дослідники вказують, що, незважаючи на широке поширення дослідницького підходу у навчанні, студенти не завжди у змозі виконати експеримент аудиторно через брак часу або матеріалів. Виконання експериментальної роботи у позанавчальний час несе додаткові ризики, особливо при роботі з небезпечними матеріалами. Використання сучасних технологій надає безпечний спосіб виконання експериментів як під керівництвом викладача, так і самостійно. Онлайн-експерименти засобами доповненої реальності та сенсорних пристроїв візуалізують для користувачів реальні дослідження і спрямовані на надання студентам можливості спостерігати й описувати роботу реальних систем при зміні їхніх параметрів та часткову заміну матеріальних ресурсів та експериментальних установок об'єктами доповненої реальності.

Т. Різов та Є. Різова [6], розглядаючи використання доповненої реальності у навчанні інженерної графіки, вводять поняття «підготовленої» та «непідготовленої» сцени (віртуального простору моделювання). Якщо програмний засіб доповненої реальності планується використовувати у «непідготовленій» сцені (як правило, поза межами аудиторії), то для визначення та відстеження її стану необхідні додаткові апаратні засоби – гіроскопи, GPS-навігатори, компаси тощо. Для аудиторного застосування доцільно «підготувати» сцену – у цьому випадку визначення положення й відстеження здійснюється за допомогою відповідних надійних чорно-білих маркерів характерної форми (квадрат або коло), що конкретизується архітектурою програмного забезпечення для їх виявлення та відстеження.

Т. П. Коделл та Д. В. Майзел [1], характеризуючи технологію доповненої реальності, вказують на простоту відображення в ній віртуальних об'єктів у порівнянні із віртуальною реальністю. Розробка об'єкту для системи доповненої реальності виконується у такий спосіб:

1) у 3D-середовищі створюється візуальна модель компоненту доповненої реальності;

2) у 2D-середовищі створюється простий маркер, що може бути швидко розпізнаний системою доповненої реальності;

3) у програмному засобі для підтримки доповненої реальності маркер пов'язується із 3D-моделлю.

При розпізнаванні маркера системою доповненої реальності на екрані пристрою із програмним засобом для підтримки доповненої реальності на зображення розпізнаного маркеру накладається відповідна йому 3D-модель. Цей процес реалізується за схемою, поданою на рис. 3.

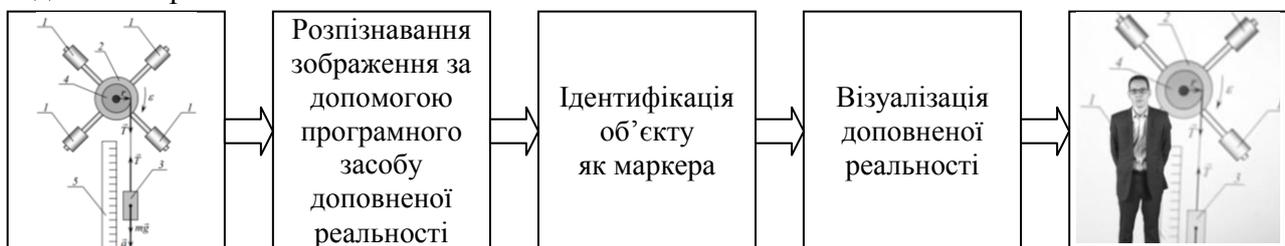

Рис. 3. Схема реалізації доповненої реальності

Використання доповненої реальності на лекційних, практичних та лабораторних заняттях полегшує розуміння студентами креслень, технічної документації та інструкцій з експлуатації. Викладачі, які використовують технологію доповненої реальності на лабораторних роботах із фізики, можуть краще пояснити студентам будову внутрішніх елементів приладів та установок, що забезпечує ефективність навчання майбутніх фахівців.

Наприклад, методичні рекомендації (рис. 4) та лабораторні установки (рис. 5) можуть бути середовищем для роботи з доповненою реальністю. Лабораторні стенди або вимірювальні прилади використовують у якості маркерів, зокрема для доповнення їх інструкціями з використання. Доповнена реальність дає сучасне вирішення завдання заохочення студентів до дослідницької діяльності та мотивує їх до експериментування.

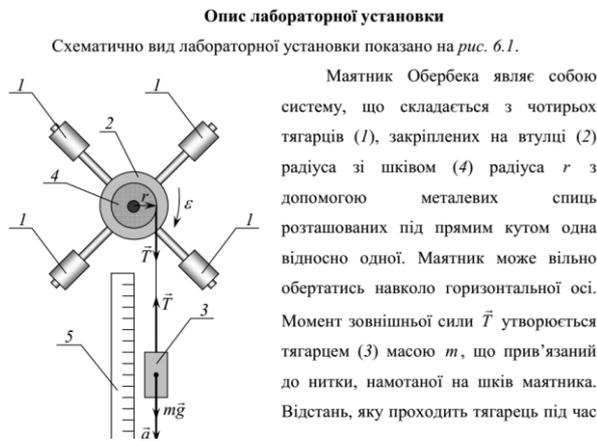 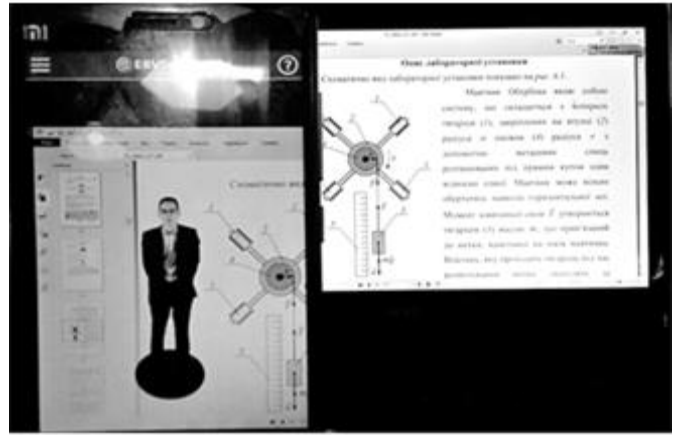

Рис. 4. Використання доповненої реальності у методичних рекомендаціях до лабораторних робіт з фізики

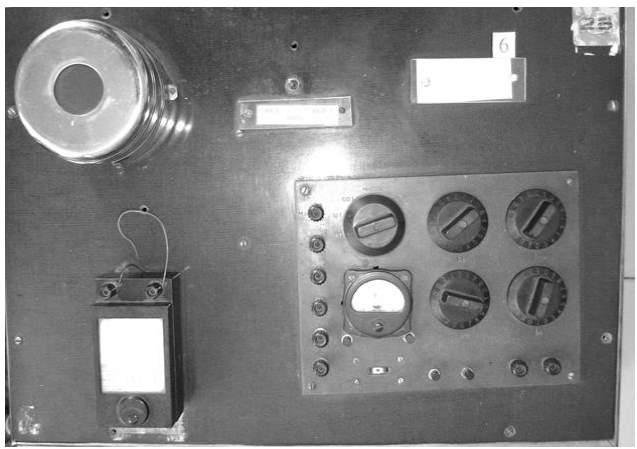 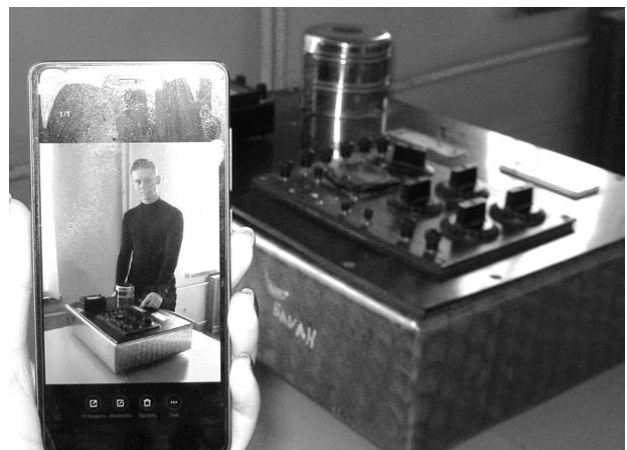

Рис. 5. Використання доповненої реальності на лабораторних роботах з фізики

Будь-який засіб доповненої реальності може бути навчальним об'єктом, якщо він є керованим та сприяє взаємодії користувача з реальними об'єктами із метою вивчення їхніх властивостей у процесі експериментального дослідження.

Робота з об'єктами доповненої реальності полягає, наприклад, у тому, щоб, використовуючи маркери як компоненти конструктора, зрозуміти принципи складання схеми електричного кола, змоделювати різні ситуації, після чого зібрати коло з реальних компонентів. У системі доповненої реальності користувач може керувати лабораторною установкою шляхом зміни положення перемикача, комбінування різних елементів, зміни положення джерела струму, його полярності тощо простим розкладанням, перекладанням та обертанням маркерів. Простота складання електричних кіл та швидка зміна конфігурації схеми дозволяє відразу ж проаналізувати результати роботи у кожній конкретній конфігурації. Такі поняття, як напрям струму, напрям обертання ротора двигуна, відкритий / закритий контур, паралельне / послідовне з'єднання легко спостерігаються і перевіряються.

Застосування засобів доповненої реальності:

– надає можливість підвищити реалістичність дослідження;

– забезпечує емоційний та пізнавальний досвід, що сприяє залученню студентів до систематичного навчання;

– надає коректні відомості про установку в процесі експериментування;

– створює нові способи подання реальних об'єктів у процесі навчання [5, с. 69-70].

Х. Мартін-Гутьєррес, Е. Гуінтерс та Д. Перес-Лопес [3] зазначають, що доповнена реальність може бути використана для спільної роботи студентів. Особливої актуальності це набуває у процесі виконання лабораторних робіт із потенційно небезпечним обладнанням, що вимагає постійного контролю діяльності студентів. Реальним лабораторним роботам передують роботи у доповненій реальності шляхом розміщення маркерів на лабораторних установках. Використовуючи маркери, студенти зможуть за допомогою мобільного пристрою візуалізувати інструкції або навчальні матеріали, необхідні для правильного використання та налаштування обладнання.

Ж.-М. Сьотат, О. Хьюг, Н. Гуаєль [2, с. 32], розглядаючи застосування доповненої реальності для активізації навчання, виділяють основні напрями її використання:

– середовища моделювання, у яких поєднуються можливості викладання, навчання, комунікації з ігровими елементами; підтримка наукових досліджень та експериментального підходу;

– перевірка моделі на адекватність; набуття технічних навичок.

**Висновки.** Використання технології доповненої реальності у мобільно орієнтованому середовищі навчання ВНЗ:

1) розширює можливості лабораторних установок, що використовуються для підготовки студентів до роботи із реальними системами;

2) робить доступними системи високої складності та вартості, які традиційно були доступні лише фахівцям;

3) надає лабораторним тренажерам інтерфейси із доповненою реальністю, що сприяє покращенню професійної підготовки;

4) мотивує студентів до експериментальної та навчально-дослідницької роботи.

**Перспективи подальших досліджень.** Розробка окремих бібліотек доповненої реальності для створення навчальних систем у мобільно орієнтованому середовищі навчання ВНЗ.

*Eugene O. Modlo[1], Yuliya V. Echkalo[2], Serhiy O. Semerikov[2], Viktoriia V. Tkachuk[2]*
[1]*Kryvyi Rig Metallurgical Institute of the National Metallurgical Academy of Ukraine*
[2]*SIHE «Kryvyi Rih National University», Kryvyi Rih, Ukraine*

**USING TECHNOLOGY OF AUGMENTED REALITY IN A MOBILE-BASED LEARNING ENVIRONMENT OF THE HIGHER EDUCATIONAL INSTITUTION**

*Research goal: to discuss the specifics of using the augmented reality technology in a mobile-based learning environment of the higher educational institution.*

*Research objectives: to determine the role and place of the technology of augmented reality in a mobile learning environment; to determine the possibilities of using the technology of augmented reality in teaching physics.*

*Object of research: a mobile-based learning environment of the higher educational institution.*

*Subject of research: augmented reality technology as a component of the mobile-based learning environment of the higher educational institution.*

*Research methods used: theoretical – analysis of scientific and methodological literature; empirical – learning and observation of the learning process.*

*Results of the research. The definition of the augmented reality concept is based on the analysis of scientific publications. It is noted that online experiments with augmented reality provide students with the opportunity to observe and describe the operation with real systems by changing their parameters, and also partially replace experimental installations with objects of augmented reality. The scheme for realizing the augmented reality is considered. The possibilities of working with augmented reality objects in teaching physics is highlighted. It is indicated that the use of the augmented reality tools allows to increase the realness of the research; provides emotional and cognitive experience, helps attract students to systematic*

*training; provides correct information about the installation in the process of experimentation; creates new ways of representing real objects in the learning process.*

**Keywords**: *technology of augmented reality, a mobile-based learning environment of the higher educational institution, teaching physics.*

**Модло Евгений Александрович[1], Ечкало Юлия Владимировна[2],
Семериков Сергей Алексеевич[2], Ткачук Виктория Васильевна[2]**
[1]*Криворожский металлургический институт Национальной металлургической академии Украины*
[2]*ГВУЗ «Криворожский национальный университет»*

## ИСПОЛЬЗОВАНИЕ ТЕХНОЛОГИИ ДОПОЛНЕННОЙ РЕАЛЬНОСТИ В МОБИЛЬНО ОРИЕНТИРОВАННОЙ СРЕДЕ ОБУЧЕНИЯ ВУЗА

*Цель исследования: обсуждение особенностей использования технологии дополненной реальности в мобильно ориентированной среде обучения вуза.*

*Задачи исследования: определить роль и место технологии дополненной реальности в мобильно ориентированной среде обучения; определить возможности использования технологии дополненной реальности в обучении физике.*

*Объект исследования: мобильно ориентированная среда обучения вуза.*

*Предмет исследования: технология дополненной реальности в мобильно ориентированной среде обучения вуза.*

*Использованные методы исследования: теоретические – анализ научно-методической литературы; эмпирические – обучение, наблюдение за учебным процессом.*

*Результаты исследования. На основе анализа научных публикаций определено понятие дополненной реальности. Отмечено, что онлайн-эксперименты средствами дополненной реальности предоставляют студентам возможность наблюдать и описывать работу реальных систем при изменении их параметров, а также частично заменить экспериментальные установки объектами дополненной реальности. Рассмотрена схема реализации дополненной реальности. Отдельно выделены возможности работы с объектами дополненной реальности в обучении физике. Указано, что применение средств дополненной реальности позволяет повысить реалистичность исследования; обеспечивает эмоциональный и познавательный опыт, способствует привлечению студентов к систематическому обучению; предоставляет корректные сведения об установке в процессе экспериментирования; создает новые способы представления реальных объектов в процессе обучения.*

**Ключевые слова**: *технология дополненной реальности, мобильно ориентированная среда обучения вуза, обучение физике.*

**ВІДОМОСТІ ПРО АВТОРІВ**

**Модло Євгеній Олександрович** – старший викладач кафедри автоматизованого управління металургійними процесами та електроприводом, Криворізький металургійний інститут Національної металургійної академії України.

*Коло наукових інтересів:* інформаційно-комунікаційні технології в освіті.

**Єчкало Юлія Володимирівна** – кандидат педагогічних наук, доцент, старший викладач кафедри фізики, ДВНЗ «Криворізький національний університет».

*Коло наукових інтересів:* інформаційно-комунікаційні технології в освіті.

**Семеріков Сергій Олексійович** – доктор педагогічних наук, професор, завідувач кафедри інженерної педагогіки та мовної підготовки, ДВНЗ «Криворізький національний університет».

*Коло наукових інтересів:* інформаційно-комунікаційні технології в освіті.

**Ткачук Вікторія Василівна** – викладач кафедри інженерної педагогіки та мовної підготовки, ДВНЗ «Криворізький національний університет».

*Коло наукових інтересів:* мобільні ІКТ, інформатичні дисципліни, професійна підготовка інженерів-педагогів.